\documentclass[smallextended]{svjour3}       
\smartqed  
\usepackage{algorithm2e}
\SetKwRepeat{Do}{do}{while}%

\def\BibTeX{{\rm B\kern-.05em{\sc i\kern-.025em b}\kern-.08em
    T\kern-.1667em\lower.7ex\hbox{E}\kern-.125emX}}
    
\usepackage[utf8]{inputenc}
 \usepackage{graphicx}
\smartqed  
\usepackage{setspace}
\usepackage{adjustbox}
\usepackage{cite}
\usepackage{amsmath,amssymb,amsfonts, pifont}
\usepackage{algorithmic}
\usepackage{hyperref}

\begin{document}
\journalname{}

\title{A Survey of Analysis Methods for Security and Safety verification in IoT Systems \\
}

\titlerunning{A Survey of Analysis Methods for Security and Safety verification in IoT Systems}        
\author{Lobna Abuserrieh \and
        Manar H. Alalfi*        
}
\institute{Lobna Abuserrieh \at
              \email{Lobna.abuserrieh@ryerson.ca} \\
           \and
           Manar H. Alalfi* \at
           \email{manar.alalfi@ryerson.ca}
}

\date{Received: date / Accepted: date}

\maketitle

\begin{abstract}
Internet of Things (IoT) has been rapidly growing in the past few years in all life disciplines. IoT provides automation and smart control to its users in different domains such as home automation, healthcare systems, automotive, and many more. Given the tremendous number of connected IoT devices, this growth leads to enormous automatic interactions among sizeable IoT apps in their environment, making IoT apps more intelligent and more enjoyable to their users. But some unforeseen interactions of IoT apps and any potential malicious behaviour can seriously cause insecure and unsafe consequences to its users, primarily non-experts, who lack the required knowledge regarding the potential impact of their IoT automation processes. In this paper, we study the problem of security and safety verification of IoT systems. We survey techniques that utilize program analysis to verify IoT applications’ security and safety properties. 
The study proposes a set of categorization and classification attributes to enhance our understanding of the research landscape in this domain. Moreover, we discuss the main challenges considered in the surveyed work and potential solutions that could be adopted to ensure the security and safety of IoT systems.
\end{abstract}

\section{Introduction}
\begin{figure*}[t!]
    \centering
    \includegraphics[width=0.99\textwidth]{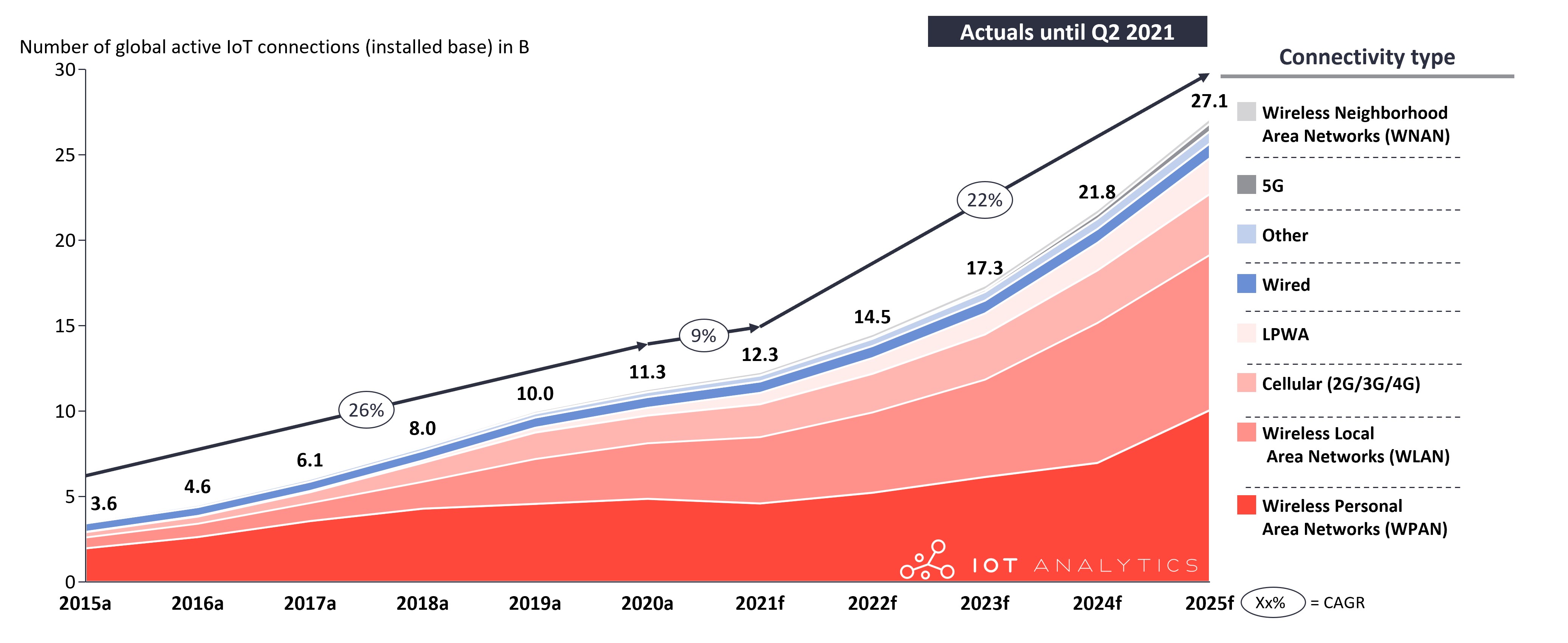}
    \caption{\textsf{Global IoT market forecast by IoTanalytics, Sept 2021} \cite{iotanalytics}}
    \label{Figure: iotana}
\end{figure*}

Internet of Things (IoT) interconnects smart objects embedded with sensors or actuators to the internet through the use of communication protocols\cite{towardIoT}. In recent years, IoT technology has been rapidly evolving; billions of IoT-enabled devices are connected and being globally used, which significantly impacts the way we control devices and the environment. Therefore, it redefines the daily routines performed and changes the way industrial tasks are done. IoT has been deployed in many different domains, e.g., healthcare, education, homes, automotive, manufacturing, entertainment, agriculture, etc. Referring to Statista, the number of connected IoT devices in 2021 jumps to over 13 billion devices, and it is expected to reach 30.9 billion devices by 2025\cite{Statista}. 70\% of automotive vehicles will be connected to the internet in  2023\cite{Statista2}. Similar numbers are announced recently by IoTanalytics as shown in Figure \ref{Figure: iotana} \cite{iotanalytics}. The figure shows the number of connected IoT devices in the past few years and an estimate of how these IoT connections would rapidly grow over time in the coming few years.

New IoT platforms are being developed, and more competitive technologies are being used in this context. Moreover, more industries are switching to use IoT technologies, leading to more complex IoT apps, more interactions among different IoT apps, and more interactions with the user. On the other hand, the growth of new IoT platforms allows heterogeneous IoT platforms to be operating in one environment. Complex interactions in heterogeneous platforms may lead to unpredicted behaviours that may affect the user or the environment, specifically, the security and safety requirements. 
For example, if the front door is opened, switch on the foyer light. And, if it is daytime, switch off the foyer light. This scenario shows two possible contradicting rules, when they are applied in one environment. But some rules could be more serious if they run in one environment. For instance,  if smoke is detected, then set the alarm on, and water sprinklers on. And if moisture is detected, then set the water valve off. These two rules are not just possibly contradicting; they may cause serious consequences if applied in one environment.
Existing IoT platforms lack the tools to verify the IoT apps' safety and security. More analysis needs to be researched and performed on IoT apps when interacting in one environment with many heterogeneous IoT platforms.
Some existing studies already covered safety and security analysis in two different scenarios: the first, analyzing a standalone IoT app in its environment. And another is considering multi-apps interaction in their running environment. Most of the current research considers three popular IoT platforms; SmartThings \cite{SmartThings}, OpenHab \cite{OpenHAB}, and IFTTT \cite{IFTTT}.
In this survey, we explore the current research in ensuring the security and safety requirements in IoT apps by considering two main search criteria. First, the program analysis techniques used in the surveyed works; secondly, how these explored studies verify the safety and security properties. Our survey explores the most recent research techniques and tools that analyze IoT apps toward safety properties, security properties, or both.  Thus, this work provides the following contributions:
\begin{itemize}
\item To the best of our knowledge, our survey is the first study that discusses different program analysis techniques applied for verifying the safety and security of IoT apps, considering apps interactions.
\item  Our survey explores  
whether the researched techniques detect, prevent, or mitigate IoT apps violations, and if the surveyed techniques can be generalized to verify the safety and security properties. Moreover, how these techniques deal with the different representations of the safety and security requirements associated with the considered IoT apps in order to be verified
\item Finally, we discuss our findings and present the research gaps in the domain of IoT safety and security verification and validation.
\end{itemize}

\section{Research Method} \label{ResearchMethod}
\textbf{Motivation and Scope}
Our motivation is to make a comprehensive study in the area of safety and security analysis for IoT Apps, and that to find  research gaps in this domain. Some related works have surveyed the safety and security in the IoT domain, but these studies did not consider all aspects related to security and safety in IoT. 

Celik et al. \cite{PACommoditySurvey,IoTPlatformsSecurityPrivacySurvey}, and Goyal et al. \cite{IoTApplicationsSecurityPrivacySurvey} explored security and privacy aspects in IoT apps. 
Jurcut et al.\cite{SecurityConsiderationSurvey}, Ammar et al. \cite{IoTSurveyonSecurityFrameworks}, and Oracevic et al. \cite{IoTSecurityServey} 
focused on security aspects in IoT. 
Paniagua et al. \cite{industrialIoTSurvey} conducted a survey to study the current state-of-art in the industrial IoT apps, while Dhanvijay et al.'s \cite{iotTechnologiesInHealthcareSurvey} explored the state-of-the-art of healthcare IoT. Finally, Ibrhim et al. \cite{ConflictsClassificationSurvey} explored and classified the conflicts in the interacting IoT apps.

Unlike the above mentioned surveys, the scope of our survey aims to explore the current research work, specifically with tools, that consider the safety issues raised from security aspects in IoT apps. Considering different comparison aspects, such as the analysis techniques, safety and security requirements specification and verification in IoT apps, how to represent these requirements, and considering the interacting environment. Therefore, we can bring light to the main gaps and problems in the analysis of safety and security for IoT Apps. 

Since our primary research scope is to explore research methods on how to verify the safety of an IoT system that is caused by a security violation, we used the following research questions to guide our search process:
\begin{enumerate}
\item \label{RQ1} What are the existing work to ensure IoT safety and security? What are the applied program analysis techniques to conduct the verification and validation? ( addressed in section \ref{sec:program analysis})
\item \label{RQ2} How are these tools and techniques identify the safety and security requirements, what are the considered factors for analysis, and how these requirements are represented as input properties to the analysis process? (addressed in section \ref{sec:SSverification})
\item \label{RQ3} What is the current gap found in the safety and security in IoT, and what are the lessons learned? (addressed in section \ref{sec:discussion})
\end{enumerate}

\textbf{Inclusion and exclusion criteria:}
Our search process was conducted using popular academic search engines, namely IEEE \cite{ieeexplore}, ACM \cite{ACM}, and ScienceDirect \cite{ScienceDirect}. We started by aligning our search with exploring one of the most related conference in Security and Safety in IoT, Usenix \cite{Usenix}, and explored some recently published work on this conference. Using Google scholar \cite{GoogleScholar}, we explored the citations of most relevant papers in this domain to find any related and recent work in the security and safety of IoT.
A combination of words was used to search the domain of IoT and Security/Safety verification together, and that to find any related work in a combined domain. We explored related papers that focused on developing smart cities, automotive, industrial and medical systems while considering security and safety validation and verification. 
An example of the pattern of queries used in our study is as follows \par

("Security "AND" Safety") AND ("IoT" OR "automation" OR" Smart cities" OR "Automotive" OR "Healthcare" OR "Medical" OR "Industry") 

Exploring the resulting papers' abstract, introduction, and conclusion to determine if the paper proposes any related approach or tool to handle any of our research questions. The final phase considered understanding each proposed work to fit our study.

\textbf{Organization:} 
We try to handle each research question in separate section, in order to make this study smooth and easily understood.
Starting with section three, a general background is presented about the related technologies; such as IoT, program analysis and the security and safety properties in IoT systems. 
The fourth section brings the light to our contribution and how the study has been developed.  Then, we handle the first research question in Section 5, that shows our study for the related surveyed papers considering the program analysis techniques. While section 6 shows the study for the literature work to discuss our second research question, that focuses on security and safety properties in IoT systems. Finally section 7 illustrate our discussion and observations resulted from comparing the surveyed tools, that discuss the last research question in our study. 
\par

\section{Background}
\subsection{Internet of Things}
Internet of Things (IoT) connects physical devices embedded with sensors or actuators to provide automated processes that could vary from simple automation like switching off the light to more complex ones, such as injecting insulin in a diabetes patient's blood. The device sensor monitors and measures some physical properties such as room temperature. In the occurrence of a specific event, these properties' values are sent to a cloud service through an IoT gateway. Such events are handled to actuate the IoT device in a predefined way.
\par IoT Platform is a multi-layer platform consisting of several software components that facilitate develop, manage and maintain IoT automation systems. Several IoT platforms are available in the market, e.g.,Samsung SmartThings \cite{SmartThings}, Apple HomeKit \cite{HomeKit}, Home Assistant \cite{HomeAssistant}, OpenHab \cite{OpenHAB}, Amazon AWS \cite{AmazonAws}, etc \cite{IoTPlatformsSecurityPrivacySurvey}. Many industries have switched recently to automate their business by taking advantage of IoT automation, e.g., Eva ICS in the industrial domain \cite{EvaICS}\cite{Patriot}, WeCon in agriculture and smart farming\cite{WeConIoT}, HealthSaaS in healthcare\cite{healthSaaS}, and many more in manufacturing, Smart cities, etc.
Trigger-action programming (TAP) is an end-user development approach that allows the IoT users to develop their own rule in the format of "If trigger- Then Action," i.e., when a rule is triggered by the specified trigger, then the associated action will be executed. TAP programming is used in IoT apps to run the action in specific events without user involvement automatically\cite{tap2021, tapSafeChain}.
One of the main issues raises when considering IoT automation is the apps interaction problem. IoT apps can perform securely and safely when each runs in isolation in its environment. But when these apps run interactively in one environment, it may cause serious security and safety issues. The interacting IoT apps could be developed or maintained using the same IoT platform or different IoT platforms. Referring back to the example of contradicting rules; When Smoke is detected, then activate the sprinkler. When moisture is detected, Switch off the water valve. If both smoke detectors and smart valve devices work interactively in the same environment, we should consider the shared environment properties. The action of one device is triggering an event associated with another device that is set up in the same environment.
\begin{table*}[t!]
    \centering
    \includegraphics[width=1\textwidth]{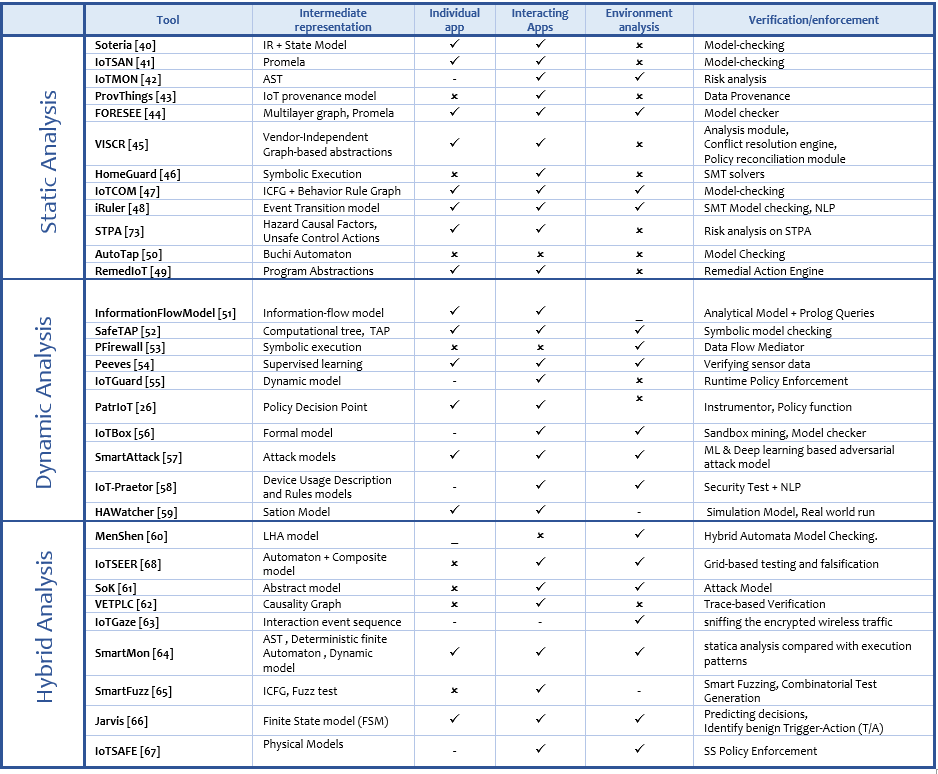}
    \caption{\textsf{Summary of Program analysis Techniques in the surveyed tools} }
    \label{tab:my_table}   
\end{table*}
\subsection{Program Analysis}
Program analysis provides analytical techniques to ensure software correctness. Different analysis techniques are adopted in the literature to ensure security, safety or both properties.
\par Static analysis conducts the program analysis by exploring the program code without the need to execute the program, in contrast to dynamic analysis, which requires the program execution to conduct the analysis\cite{StaticDynamicHybridComparison,DynamicAnalysisSurvey}. Literature studies adopt different program analysis techniques to analyze and verify IoT apps properties. 
\subsubsection{Static Analysis}
\begin{itemize}
    \item \textit{Abstract Syntax Representation:} helps extract the possible set of states of the program. The code is analyzed and represented using an Abstract Syntax Tree (AST).
    \item \textit{Symbolic Execution} is one of the program analysis techniques and it is well-adopted in both static and dynamic analysis. Symbolic execution uses some mathematical expressions using symbols for the program inputs and specifies which execution path will be taken. Typically Satisfiability modulo theories (SMT) solvers are used for Symbolic execution\cite{SMTSolvers}. 
    \item \textit{Control Flow Graph:} building a graph to represent each possible traversed path in the analyzed program\cite{modelChecking} .
    \item \textit{Model checking:} a technique to explore a finite set of possible program states and check whether it meets a given specification.
    
    \item \textit{Data-driven analysis:} a method to find a possible set of specific values calculated out from tracing the program codes
\end{itemize}
\subsubsection{Dynamic Analysis}
\begin{itemize}
    \item  \textit{Symbolic Execution }, the dynamic symbolic execution aims to change the decision conditions in the program under check to build new execution paths and check if the newly generated paths are reachable \cite{SMTSolvers, symbolicExecutionSurvey}.
    \item 
    \textit{Tracing and instrumentation: } aims to analyze executable data and verify logs to diagnose errors and trace logs; as a technique to measure apps performance.
    \item \textit{Input generation for testing:} Generating input data to run apps testing. Some AI and ML techniques could be utilized to improve this process.
    \item \textit{Security test:} a testing technique that monitors and assesses the app's security to detect security problems.
    \item \textit{Run-time enforcement:} a technique used to monitor the app's behaviour in its execution. When a predefined property is violated, the execution is blocked.
\end{itemize}
\subsubsection{Hybrid Analysis}
Hybrid program analysis takes advantage of combining both static and dynamic techniques in order to improve the quality of the analysis results.
\subsection{Security and Safety }
Many recent research studies highlight the challenges of securing IoT systems, especially in open internet networks and heterogeneous technologies. Therefore, developing large-scale IoT systems will increase security threats\cite{SecuringIoT,SecuringIoTRoadAhead}.  The increasing number of connected IoT devices involved in daily life routines makes security breaches critical to the users' safety. The literature shows the close relationship between the safety and security of IoT systems. Safety ensures that deploying IoT systems does not impact its users' and environment safety while ensuring that the IoT environment does not cause any security breaches to the IoT system\cite{IoTSafetyStateofArt}. One exciting example we like to mention here is when hackers attempted to remotely kill a Jeep Cherokee system, which was a big motivation to dig more into this domain. The hackers have shown how the security breach in an automotive IoT system could seriously impact its users' safety\cite{JeepCherooke}.

\section{Program Analysis Techniques}
\label{sec:program analysis}
To answer \textit{RQ\ref{RQ1}} mentioned in the research method in section \ref{ResearchMethod}, this study considers exploring program analysis techniques applied to ensure that the IoT systems meet the specified security and safety requirements when deployed in their running environment. This section focuses on how the related published tools analyze the IoT apps and the applied analysis pattern in their logic, compares some tools and highlights if they use other techniques to improve the tool effectiveness.
The surveyed tools are categorized into groups based on their analysis type; static, dynamic, or hybrid. Each group is shown in Table \ref{tab:my_table} with the main factors we are surveying. In static analysis, the IoT app execution is not required to be analyzed, whereas dynamic analysis requires the app to run to analyze it. In comparison, the hybrid analysis combines both static and dynamic analyses to analyze the application. The following subsections cover the three groups in the surveyed works.

\subsection{Static Analysis}
Different static analysis techniques were utilized in the surveyed tools. 
Soteria translates the groovy code of a given IoT app into Intermediate Representation (IR) and then to State-machine models. The NuSMV model checker is used to verify its conformance towards the specified security and safety properties\cite{Soteria}. 

IoTSAN analyzes IoT apps interactions by constructing Dependency Graphs (DG). And by using Bandera Tool Set, IoTSAN feeds the generated Promela models into the SPIN model checker to detect any potential safety violations\cite{Iotsan}. 

IoTMON proposes exploring all possible physical interactions among IoT apps by running intra-app interaction analysis \cite{IoTMoNn}. IoTMon tool built on three main steps: (i) Model the safety properties using General Policy Model. (ii) Perform the intra-app interaction analysis by extracting the Abstract Syntax Tree (AST) of the given IoT apps. (iii)Trace the control flow to capture triggers and actions. IoTMon is not designed to detect policies’ violation in real environments \cite{IoTMoNn}.

ProvThings provides a complete history of IoT system interactions, which may guide the analysis to the potential malicious behaviour. ProvThings uses provenance-based tracing to generate a data provenance graph to perform the analysis \cite{ProvThings}. 

ForeSEE validates an IoT system's security properties by building a multi-layer IoT hypothesis graph and performing model checking using the SPIN model checker\cite{Foresee}. 

VISCR translates IoT groovy programs into vendor-independent graph-based abstractions, then it validates these generated abstractions and detects potential conflicts or violations \cite{viscr}.

HomeGuard extracts rules automation from IoT apps by applying symbolic execution and identifies Cross-App Interference (CAI) threats by encoding each of these threats type into Satisfiability Modulo Theory \cite{HomeGuard}. i.e., HomeGuard analyzes each pair of automation rules and its specified configuration towards SMT. The prototype provides symbolic execution to statically analyze the Abstract syntax Tree to extract the rules of IoT apps and conducts an app instrumentation approach to fetch the users' configurations. 
Moreover, HomeGuard provides a risk ranking technique that assesses the risk seriousness as high, medium or low, based on the impact of rule execution, its functionality, and criticality of the device security. Based on the work evaluation, they were able to identify 663 CIA violations in 146 SmartThings real apps.
IoTCOM translates the groovy code of an IoT app to interprocedural control flow graphs (ICFG) and then to Behavioral Rule Graph (BRG) to run model checking \cite{IoTcomm}. 
iRuler analyzes IoT app inter-rules security violations by building Rule Representations (RR) models along with the information flow model\cite{iruler}. 
RemedIoT uses programming abstractions, actuation graphs, actuation modules, and policy grammar \cite{Remediot}.
AutoTap performs Rule-based Analysis. AutoTap allows the users to specify their desired properties. It uses Buchi Automaton to translate these properties into linear temporal logic (LTL) and ensure it follows the TAP rules\cite{Autotap}.

\subsection{Dynamic Analysis}
The literature introduces different dynamic analysis techniques that have been utilized in verifying the security and safety of IoT systems.
Users can create their daily recipes, IFTTT applets in 2016; these recipes can be published on social media and captured by others. Surbatovich et al. \cite{infoFlowModel} proposed an information flow model to analyze such IFTTT recipes to detect any security or privacy risks that could be harmful if used by attackers. The model and the dataset are coded using Prolog facts, and then the analysis is run using Prolog queries. 
The analysis assigns labels to IFTTT triggers and actions based on the description associated with each recipe.  Then, it finds the direct links between any two existing recipes and determines the existence of any physical connections, which should have been previously tracked when triggers and actions were labelled. The evaluation shows 49.9\% of the dataset recipes were unsafe due to secrecy or integrity violations.
\par SafeTap tool was developed to allow TAP users to check their apps toward their desired behaviours \cite{Safetap}. SafeTap was proposed to analyze TAP apps on performing Symbolic Model Checking SMC. In comparison, SafeTap\( \Delta \) has improved SafeTap by extending the SMC to deal with the updated TAP rules to detect any newly introduced violations. SafetTap takes IoT apps rules as a labelled-transition system and the desired behaviour by representing the properties in computational Tree Language ( CTL). After that, it monitors the current state of the running IoT apps to perform its analysis. Model checking verifies if the modelled system satisfies the set of specified properties. SafeTap\( \Delta \) is extending SafeTap by applying incremental analysis for any updated rules and providing feedback to the user in case of any violations. Their study shows that SafetTap helped its users to perform better to identify the properties’ violations. 
\par PFirewall system monitors the data between the IoT device and the hub, and filters the data based on specified privacy policies\cite{Pfirewall}. Therefore, the system's primary role is to ensure privacy protection by monitoring the data flow without changing the existing IoT apps or platforms. PFirewall overcomes the hub encryption challenge by applying the man-in-middle technique; it connects with all existing devices as a hub. It then establishes a set of the same number of virtual devices to be paired with the actual hub. Applying PFirewall ensures data minimization and enforces users' privacy. It runs on both SmartThings and OpenHab platforms, and its evaluation shows a 97\% reduction in the sent data.
\par Birnbach et al. have introduced Peeves to predict event spoofing in smart homes, which is an event that could trigger an action, but it did not physically happen due to a sensor fault or attack attempt\cite{Peeves}. Peeves system uses data from the connected sensors to automatically verify the physical events. Peeves makes use of supervised learning to utilize events signatures and sensors' physical signatures. The evaluation was run over two weeks in an actual real-world smart home environment with 48 connected sensors and 22 event types with 100\% detection rate\cite{Peeves}.
\par IoTGuard is a dynamic policy enforcement tool that supports safety and security violations detection and enforcement in IoT apps\cite{Iotguard}. IoTGuard provides three main components to dynamically analyzing the IoT app. First, it applies the instrumentation component to collect the app's runtime-related information. Then, IoTGuard observes the execution behaviour to build the dynamic execution model of the IoT app. Last, it specifies the policies to be enforced on the execution model. This process detects violations on an individual IoT app or a group of IoT apps interacting in one environment. The app's users can allow the violations to be executed or let IoTGuard block these violations. IoTGuard is evaluated on SmartThings apps and IFTTT applets.
\par Another dynamic detection approach was proposed by Yahyazadeh et al.\cite{Patriot},  PATRIoT, for Policy violation detection. PATRIoT main idea is to monitor the IoT app during its execution, and when any specified policy is violated, PATRIoT will block the app's action. PatrIoT targets three IoT platforms, EVA ICS for industrial automation, and SmartThings and OpenHAB for home automation. PATRIoT evaluation was tested using 33 policies in 122 SmarThings apps, 10 policies in 20 OpenHab rules, and 6 policies in 8 EVA ICS automation sets.

\par Another dynamic technique, IoTBox, uses sandbox mining to prevent undesired behaviour \cite{IoTBox}. IoTBox observes the current behaviours of smart homes. If any behaviour is not previously explored, e.g., the app is updated, or malicious code exists, IoTBox prevents the behaviour if it violates its rules, and this behaviour is reported. The user may revise the action and allow the behaviour if it is trusted. IoTBox adds this behaviour to the sandbox to be relearned for future behaviours. IoTBox utilizes formal models and model checking to ensure the conformance of the expected behaviour and the actual execution. If not, IoTBox warns the user.
\par
Yu and Chen \cite{Smartattack} has introduced the SmartAttack framework that relies on deep learning and machine learning to design and build attack models for IoT-based apps. Users can configure the attack model in the best suitable to their needs. It enables evaluating the security and privacy properties in smart home automation systems. 
SmartAttack has been evaluated using two datasets, UNSW Sydney smart home dataset and a mock smart home dataset, SmartFIU. 
\par
IoT-Praetor system detects the undesired behaviour of an IoT app by using Device Usage Description(DUD) and generating the behaviour model, Device usage rules (DUR), including the app's interactions and communications. It dynamically collects the app behavior and designs a behavioural rule engine to expand its flexibility in analyzing real-time behaviours. IoT-Praetor was evaluated using SmartThings apps and IFTTT applets, and it has shown 94.5\% detection rate of malicious interactions and over 98\% in detecting malicious communications\cite{IoTPraetor} .
\par
HAWatcher approach is an anomaly detection system for SmartThings  IoT apps\cite{HAWatcher}. 
HAWatcher first builds a simulation model for normal home automation behaviour by capturing both events logs and system semantics, and it observes the automation process at runtime. This process enables HAWatcher to explore the differences and inconsistencies between the real world and the simulation model to detect violations. HAWatcher was evaluated using four Smartthings datasets against 62 different violations.
\subsection{Hybrid Analysis}
Menshen is a framework to automatically build and check the Linear Hybrid Automata (LHA) model of the IoT system\cite{MenShen}. The checking process of Menshen is to detect violations and propose fixing suggestions to its users. MenShen conducts its verification process by implementing three main steps: (i) Linear Hybrid Automata Automatic Modeling to generate LHA models, (ii) Analyzing LHA model reachability to check the system against good and bad states based on path-oriented checking, (iii) and lastly Fix suggestions synthesis to provide fixing suggestions if the verification fails.  The evaluation was conducted on sets of real HA-IoT 46 devices and 65 rules, with results showing a run time of 10 seconds to detect violations and provide fix suggestions.
\par
SOK proposes an abstract graph model to represent IoT deployments and their topologies. SOK uses the security properties, attack vectors, mitigations, and stakeholders, to build a systematization approach that uses the proposed abstract model and then assesses the security properties. SOK has been evaluated by observing 45 devices and found a total of 84 running services and 39 issues related to those running services\cite{SOK}.
VETPLC is a framework for safety vetting in Industrial Controller systems (ICS), mainly to detect Safety violations caused by programs faults or attack attempts\cite{VETPLC}. What bought our attention to include this study is two main reasons. First, we recently see a vast revolution in applying IoT in the industrial domain, and some IoT platforms are specially dedicated to handling industrial automation, such as Eve ICS. Second, analyzing systems to detect safety violations using a causality graph differed from other studies we have explored. VETPLC is used for automatic safety vetting by building timed event causality graphs to explore causal relations among events in the code. Then VETPLC mines temporal invariants from data traces collected in Industrial Control System (ICS) to verify any temporal dependencies. 
\par
IoTGaze framework considers detecting security vulnerabilities in IoT systems by analyzing the connected wireless traffic\cite{IotGaze}. IoTGaze first identifies events sequential interactions between apps and devices. IoTGaze uses all IoT context by extracting user interfaces, app descriptions, and specified expectations. Therefore, IotGaze can detect anomalies when the wireless context analysis does not match the IoT context expectations.  The evaluation is conducted using the Samsung SmartThings framework.
\par
SmartMon is a network monitoring platform that monitors network events and verifies the behaviours of smart home devices by collecting and analyzing network traffic \cite{SmartMon}. SmartMon is intended to detect security violations by analyzing IoT apps statically by analyzing the app source code and building its Abstract Syntax Tree, and dynamically by monitoring the app at runtime and device states changes. Then SmartMon applies a match module to verify the device state change happens correctly to confirm the trigger-action chain. SmartMon evaluation reached more than 95\% of violations detection, and it was evaluated in an environment where multiple apps were executed simultaneously with some dependencies.
SmartFuzz Is an automated test generation approach based on static analysis by constructing inter-procedural control flow graph (ICFG) using ContexIoT, and dynamic analysis by extracting the possible values of IoT apps attributes to improve the test coverage. SmartFuzz then uses Selenium-based and combinatorial testing for test cases generation. Finally, SmartFuzz produces test reports after running its testing on the SmartApp using the generated test cases. Evaluating smartFuzz was only examined on 60 SmartThings apps and IoTBench  \cite{SmartFuzz}.
\par
Jarvis is a Reinforcement Learning technique for predicting the optimal and safe IoT systems decisions\cite{Jarvis}. Jarvis models the simulated IoT environment based on the states and actions.  And then, agents in RL explore the optimal actions based on users' specifications or goals. Therefore, by utilizing agents in reinforcement learning, Jarvis can predict the optimal behaviour and actions. Deep Q learning is used to decide the best optimal actions concerning timing period and user goals, while agents are trained using a deep neural network (DNN).
\par
IoTSafe is an approach for dynamic security and safety enforcement in an IoT environment\cite{Iotsafe}. The proposed IoTSafe is built on analyzing physical interaction, including temporal, in multi-app IoT systems by utilizing static and dynamic testing techniques. IoTSafe analyzes SmartThings apps. IoTSafe evaluation was run on 130 potential physical interactions, and it has successfully identified 39 real  physical interactions
\par
IoTSEER is proposed to detect undesired security behaviour caused by potential physical interaction\cite{IoTSEER}. IoTSEER explores the IoT app source code and translates it to a hybrid automaton to define the physical interaction and builds a new automaton for the joint behaviour representing the interacting IoT apps. Then, IoTSEER checks whether the actual IoT system conforms to the specified policies and provides a violation cause report to guide the users. The evaluation of IoTSeer was conducted in a real home automation environment with 37 installed apps using 13 actuators and six sensors.

\section{IoT apps Security and Safety Verification}
\label{sec:SSverification}
Verifying security and safety in IoT apps doesn't only analyze the system, but it also needs to have a set of properties input to the verification engine. Hence, we should consider the IoT analysis process itself and how the properties are represented and fed to the analysis process. Some of the literature tools don't consider defining the security and safety properties, and rely on analyzing the system towards well-known issues; e.g., security attacks\cite{Smartattack, SmartMon, SOK, Hybridiagnostics}, wireless network security, shared space problem, etc. Some of the explored tools apply their approach by comparing the system execution obtained from dynamic testing techniques, with the system specification that is analyzed statically\cite{SmartFuzz, Peeves, HAWatcher}.

To address \textit{RQ\ref{RQ2}}, mentioned in section \ref{ResearchMethod}, we first discuss and categorize various security and safety properties used by researchers, then we discuss the various representations used to describe the security/safety properties to be input to the verification engine, and finally, we demonstrate the different approaches used to verify the  IoT app against these described properties. Table \ref{tab:SS_ToolsTable} summarize the explored techniques, the domain of properties they consider, and the tools relation to Security and Safety properties.
\begin{table*}
    \centering
    \includegraphics[width=1\textwidth]{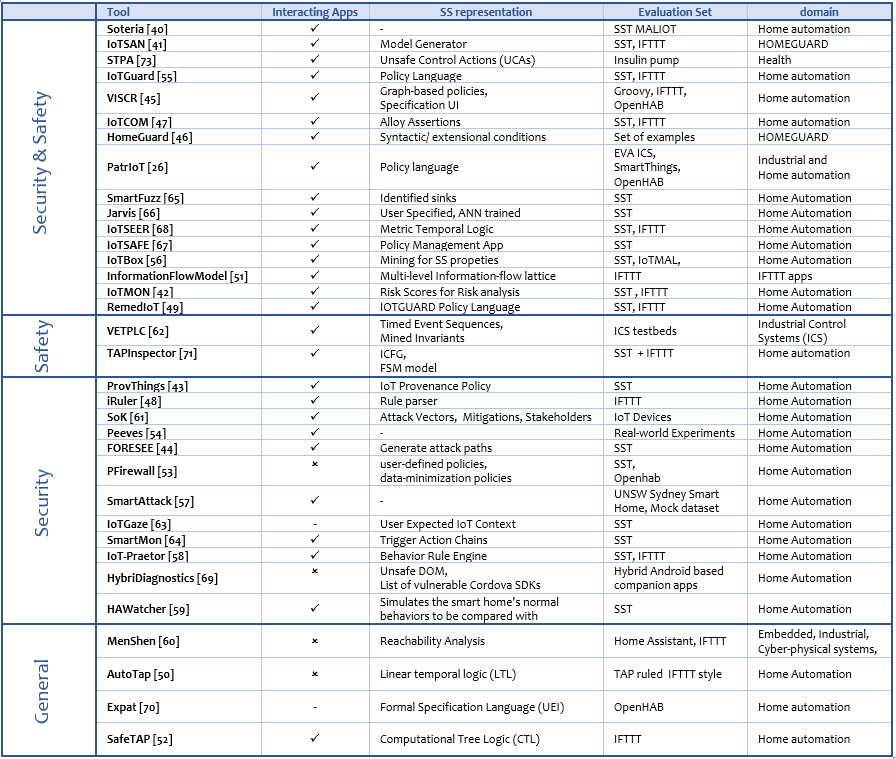}
    \caption{\textsf{Summary of the surveyed tools analyzing IoT security and safety} }
    \label{tab:SS_ToolsTable}   
\end{table*}

\subsection{\textbf{IoT app Security and Safety properties}}
\subsubsection*{\textbf{General properties}}
    \begin{enumerate}
         \item 	Individual: one app with some violations
            \begin{enumerate}
                 \item Conflicting values in one control path; e.g., Motion detector, can't change the value of switch attribute to on and off in the same control flow branch
                 \item An attribute value changes many times in the same path to the same value,
                    Example: Motion detector, must not give the same value of switch attribute multiple times in the same control flow branch.
   
            \end{enumerate}
        \item Interacting Apps
            \begin{enumerate}
                \item Race condition in many event handlers: two or more handlers must not change one attribute value.  Example: door unlocked changes the switch attribute value to on, while sunrise event changes the switch to off. Even if it is the same value, we should consider it.
                 \item 	Simultaneous events/action: two or more handlers are triggered at the same time; One can finish before the other one. This mean the environment values could be changed by the first one while the second is still working on the previous environment value. Example:                \[ Smoke Detected \rightarrow Alarm On \: \& \: SprinklersOn,  \] \[SprinklersOn \rightarrow put\:out\:the\:fire\:and\:smoke. \] \textrm{While the alarm is still on.}
            \end{enumerate} 
        \item Dependent Apps
            \begin{enumerate}
                \item Cyclic events/loops; When an Event triggers a handler that triggers the first event. E.g. \[DoorLocked \rightarrow SwitchOn, \]  \[SwitchOn \rightarrow  DoorLocked.\]
            \end{enumerate}
    \end{enumerate}
\par
\subsubsection*{\textbf{App Specific properties}}
\begin{enumerate}
    \item Professional/ expert defined: in which developers or customers specify the requirements based on the business needs.
    \item User defined: in this type of properties, the IoT app allows app's end users to specify and customize their properties based on their own needs.
\end{enumerate}
\subsection{\textbf{Environment Impact on IoT apps interactions}}
\begin{figure}[t!]
    \centering
    \includegraphics[width=0.7\textwidth]{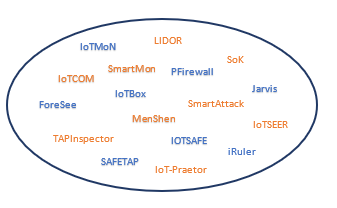}
    \caption{\textsf{Techniques that consider IoT Environment}}
    \label{figure: envTools}
\end{figure}

Some of the techniques we explored consider the environmental factors surrounding the IoT apps in where they work or interact. At the same time, many other techniques do not consider the environmental factors. Figure \ref{figure: envTools} shows the techniques that consider physical environment or channels.

\subsection{\textbf{Representing IoT Security and Safety properties}}
Different representations for Security and Safety properties have been employed when these properties were given as input to the analysis process. Table \ref{tab:my_SStable} shows the literature tools and how Security and Safety properties are represented in each tool, along with the verification type used.
\clearpage
\begin{table*}[!ht]
    \centering
    \includegraphics[width=0.99\textwidth]{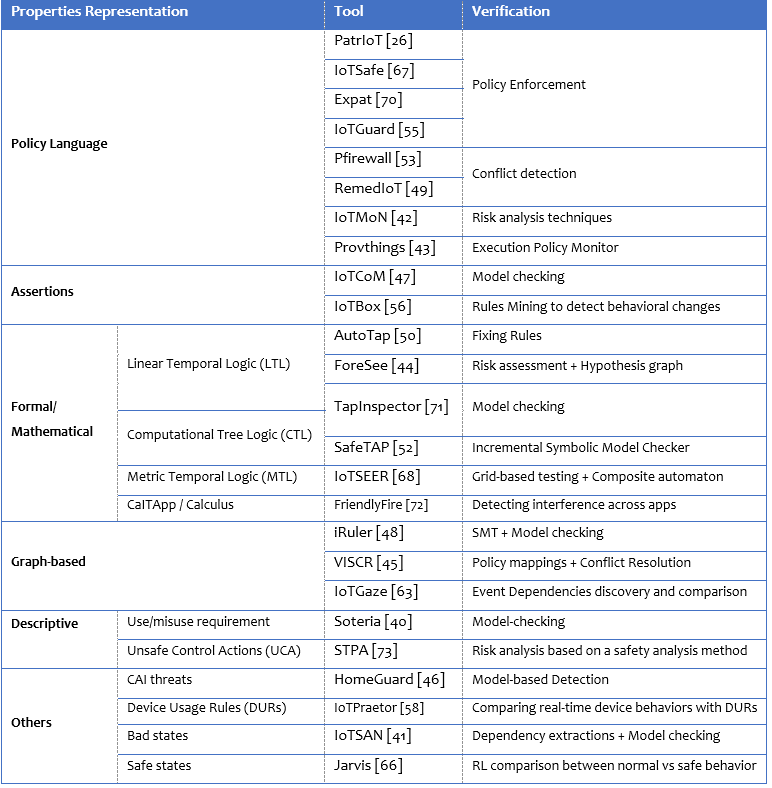}
    \caption{\textsf{Summary of Requirements Representation \& verification in the surveyed tools} }
    \label{tab:my_SStable}   
\end{table*}
			
\subsubsection{Model-based Representation} 
\begin{figure}[t!]
    \centering
    \includegraphics[width=0.9\textwidth]{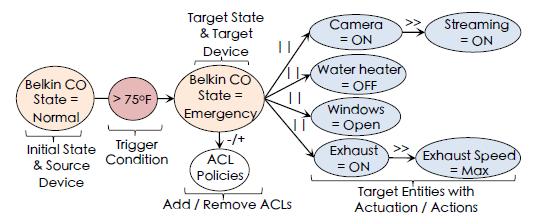}
    \caption{\textsf{VISCR Graph-based Policy \cite{viscr} }}
    \label{fig:viscr}
\end{figure}

The properties representation used in these tools is based on graphs models. Most of the tools use this kind of representation to facilitate the model-checking process for analysis \cite{viscr,iruler,IotGaze}. 
iRuler represents inter-rule vulnerabilities as flow graphs \cite{iruler}. IoTGaze utilizes the dependency graphs to conclude security properties\cite{IotGaze}. VISCR generates dynamic graph based along with Access Control List (ACL) policies for the specified properties\cite{viscr}.  VISCR can also represents the policies using finite state automation. For example, Figure \ref{fig:viscr} shows an example for a state where the IoT device detects heat and the trigger condition occurs, heat $ > $  75\textdegree  F. Thus, a state of emergency is applied as an action; turning both camera and exhaust On, turning water heater off and windows open. the symbol $ \| $ is used to indicate that these actions are parallel, while the symbol $ >> $ is used to indicate the sequential actions.

\subsubsection{Policy Language/specification} 
\begin{figure}[!ht]
    \centering
    \includegraphics[width=0.7\textwidth]{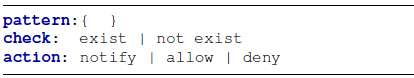}
    \caption{\textsf{ProvThings Policy Language Description\cite{ProvThings}.}}
    \label{fig:provthingsf}
\end{figure}

Some existing studies propose policy language in order to verify the system correctness against its properties. For instance, ProvThings deploys a Policy Monitor as one main component of its verification server. The Policy Monitor component ensures that the execution conforms to the defined policies; when the action allows a specific function to perform, the control executes the action. Otherwise, the control code goes to the following statement and skips the action's function execution. 
All of PatrIoT \cite{Patriot}, IoTSafe \cite{Iotsafe}, Expat \cite{Expat}, and IoTGuard \cite{Iotguard} follow the same semantic in enforcing the predefined policies. While both PFireWall and RemedIoT implement their own conflict detection engine to verify the policies conformance \cite{Pfirewall,Remediot}. Whereas IoTMoN relies on risk analysis techniques to ensure the policies well met \cite{IoTMoNn}. ProvThings provides a policy monitor to keep ensuring that the system provenance flow still conforms to the specified policies\cite{ProvThings}.
For example, ProvThings policy description is shown in Figure \ref{fig:provthingsf}. The figure shows three parts: The policy pattern to search for while the system is running, whether the need is to check the pattern existence or absence from the system execution, and if the pattern matched to exist or not, what is the necessary action to be enforced in this situation.

    Some more tools use Policy Language to represent the Safety or Security properties, e.g., Expat\cite{Expat}, IoTGuard\cite{Iotguard}, PatrIoT\cite{Patriot}, RemedIoT\cite{Remediot}.
\subsubsection{Properties as Formal Assertions}
\begin{figure}[!ht]
    \centering
    \includegraphics[width=0.8\textwidth]{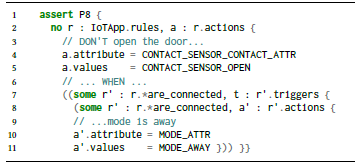}
    \caption{\textsf{IoTCoM Alloy Assertion \cite{IoTcomm}.}}
    \label{fig:iotcomFF}
\end{figure}

Some of the explored tools describe the properties as assertions in order to verify the IoT app against these properties. For instance, IoTCoM translates the specified properties to Alloy Assertions, then they were used by the Alloy model checker in verifying the home model\cite{IoTcomm}. IoTBox uses the same policies representation and deployed Alloy Assertions as the system defined policies\cite{IoTBox}. 
Mohannad et al. \cite{IoTcomm} provided the assertion representation in Fig \ref{fig:iotcomFF} to describe the property "DON'T unlock door WHEN location mode is Away." \\
\clearpage
\subsubsection{Formal and Mathematical representations}
\begin{figure}[!h]
    \centering
    \includegraphics[width=0.6\textwidth]{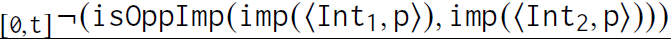}
    \caption{\textsf{IOTSEER Policy using MTL \cite{IoTSEER}.}}
    \label{fig:iotseerMTL}
\end{figure}

IoTSEER uses Metric Temporal Logic (MTL) to describe physical interactions with undesired consequences on the system behaviour, which then can be used by Grid-based testing to validate the IoT system \cite{IoTSEER}. MTL representation are formulas used to describe the specified properties. For instance, IoTSEER describes an example policy \textit{"Multiple intended physical channels must not influence a sensor event in opposite ways"\cite{IoTSEER}} using the MTL representation in Figure \ref{fig:iotseerMTL}. The MTL representation uses \textit{Int} for intended influence,  \textit{unInt} for unintended influence, and \textit{isOppImp} for opposing influence from different actions.
Similarly, temporal logic representation is used by SafeTAP\cite{Safetap}, AutoTap\cite{Autotap} and TapInspector \cite{Tapinspector}.
ForeSee correctness properties specified as Linear Temporal Logic (LTL) formulas \cite{Foresee}. While Jarvis represents safe states and transitions in defined formulas \cite{Jarvis}. On the other hand, FriendlyFire uses calculus in representing the properties\cite{FriendlyFire}.
\subsubsection{ Descriptive representation}
STPA uses Unsafe Control Actions (UCA) to describe undesired safety behaviour\cite{stpa}. STPA relies on three main descriptions in their analysis: Accidents which are Undesirable events, Hazards describing the conditions that leads to the accidents and finally the Safety constraints (SC) Constraints to prevent hazards. All the factors are described by text along with a probability value which is given to the accidents. The example in Table \ref{tab:STPA} shows the basic tabular descriptive representation of the used safety properties in STPA.
\begin{table}[!ht]
    \centering
    \includegraphics[width=0.8\textwidth]{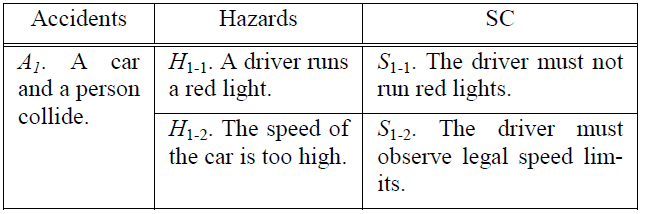}
    \caption{\textsf{STPA Accidents, Hazards, and Safety constraint\cite{stpa}} }
    \label{tab:STPA}   
\end{table}
\\
Soteria  describes the properties in text, but it uses use/misuse requirement engineering technique to formalize it to be used in model checking.
\par
    \begin{figure}[ht!]
    \centering
    \includegraphics[width=0.9\textwidth]{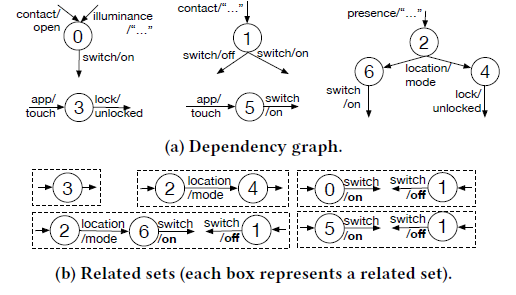}
    \caption{\textsf{IoTSan Dependency Extraction \cite{Iotsan}.}}
    \label{fig:iotSanP}
\end{figure}
\begin{figure*}[ht!]
    \centering
    \includegraphics[width=0.99\textwidth]{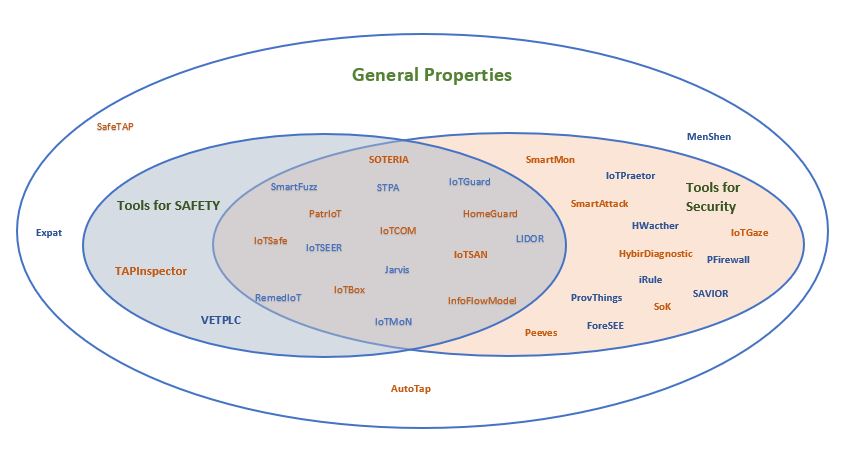}
    \caption{\textsf{Tools for Analyzing IoT Security, Safety, or General properties}}
    \label{fig:ds}
\end{figure*} 
\subsubsection{Other representations }
some tools deploy dependency extractions of bad states as violations like IoTSan \cite{Iotsan}.  Figure \ref{fig:iotSanP} shows an example how to extract the dependency from state model in order to extract bad states. IoTSan identifies the safety properties violations as bad physical states to be used in model checking and the modelled system. 
    
 On the other hand, Jarvis relies on safe states to be able to run Reinforcement Learning (RL) to compare the the normal behavior vs the safe behavior \cite{Jarvis}. IoTPraetor uses the Device Usage Rules in order to compare the device behavior with these predefined rules\cite{IoTPraetor}.
\subsection{Different verification domains}
When analyzing IoT apps for Security and Safety requirements, we can classify  tools presented in the literature into four main categories:
\begin{itemize}
    \item The first category was the first target, those tools target analyzing IoT apps towards their Security and Safety requirements. E.g., Soteria\cite{Soteria}, IoTCom \cite{IoTcomm}, IoTSan \cite{Iotsan}, HomeGuard\cite{HomeGuard}, etc.
    \item The second set of tools focuses on analyzing IoT apps for Safety properties only. E.g., TapInspector\cite{Tapinspector}, VetPLC\cite{VETPLC}.
    \item Third is the set that focuses on analyzing IoT apps toward their conformance to security properties. For instance, SmartMon\cite{SmartMon}, IoTGaze\cite{IotGaze}, ForeSeee\cite{Foresee}, Savoir \cite{Savior}.
    
    \item 	The last set analyzes IoT apps towards some general requirements, which can be customized to Security and Safety. For example, Expat\cite{Expat}, SafeTap\cite{Safetap}, MenShen\cite{MenShen} and AutoTap\cite{Autotap}.
\end{itemize}
Figure \ref{fig:ds} summarizes the four categories showing tools in the literature that analyze Security and Safety properties.

\section{Discussion and Lessons learned}
\label{sec:discussion}

This section discuss the last  research question \textit{RQ\ref{RQ3}} presented in section \ref{ResearchMethod}, highlighting the main discussion points, observations and gaps found while exploring the literature work. 
\subsection{Optimization Techniques}
Some literature tools improve their analysis techniques by utilizing different techniques, e.g., AI and ML techniques, or even combining both static and dynamic analysis to produce better results.
IoTGaze\cite{IotGaze}, SmartMon\cite{SmartMon}, IotSafe\cite{Iotsafe}, IotSeer\cite{IoTSEER}, and more tools , shown in Table \ref{tab:my_table}, applied Hybrid analysis by combining both static and dynamic analysis to achieve better results.
Some Literature tools consider applying different Artificial Intelligence (AI) and Machine Learning  (ML) techniques to improve the analysis of IoT systems. For example,  Jarvis \cite{Jarvis} uses Reinforcement Learning, Deep Q learning network, along with deep neural network to train the agent in order to improve the detection and achieve the best quality of the results. IoTMon\cite{IoTMoNn}, iRuler\cite{iruler}, and IoT-Praetor\cite{IoTPraetor}, uses Natural Language Processing (NLP) in their IoT apps analysis.
Applying such techniques helps improve the quality of the results and enhances the analysis process.
\subsection{Tools Availability }
Some tools are available to be used publicly; for instance, IotSan\cite{Iotsan}, SmartAttack\cite{Smartattack}, IoTCom\cite{IoTcomm}, SoK\cite{SOK}. But many more are not available for public, e.g. Soteria\cite{Soteria}.

\subsection{Heterogeneous Platforms}
Recently we can see a rapid growth of new IoT platforms, which can introduce a new challenge; How to deal with heterogeneous IoT platforms in one operating environment? This issue can be more explored. Some of the literature tools could be improved to conduct the analysis process in one environment with different IoT apps from various IoT platforms.
\subsection{Early Developmental Consideration}
A few tools proposed risk analysis by applying risk scores to the IoT actions and considering them while performing the analysis. This approach could improve the analysis process, especially when it's considered in the early IoT developmental stages;  it helps formalize the critical properties needed to be considered in analyzing the security and safety of IoT systems.

\subsection{Further Discussion and Observations }
Some literature approaches don't consider formalizing Security and Safety properties to be fully automated when conducting the analysis process. We noticed the focus is more on analyzing the IoT system by understanding its components while feeding the system with the Security and Safety properties was not formally utilized in many of the literature tools.  Some tools use a text-based description for security and safety properties and conduct manual conformance to IoT system analysis toward these properties. Some other tools have paid attention to this issue, but their analysis technique was based on transforming the IoT system to many representations, which affect the general analysis performance. The lack of Security and Safety standards in IoT systems also adds difficulties to this process.
\section{Related Work}
The swift growth of the IoT automation systems makes the research area in this domain rapidly developed. Many new tools explored in this study are just released, e.g. IoTBox\cite{IoTBox}, HAWatcher\cite{HAWatcher}, IoTSEER\cite{IoTSEER}, IoTSafe\cite{Iotsafe}, HybridDiagnostic\cite{Hybridiagnostics}, Tapinspector\cite{Tapinspector}, etc. Therefore, many previous survey studies do not include these newly released tools in their comparison. Some related surveys focus on a specific property, analysis type or improvement technique, while our research combines all and discusses some related issues and challenges in this regard. \par
Oracevic et al.\cite{IoTSecurityServey} explore and focus only on the security of IoT systems and associated challenges and open issues. This survey was published in 2017 many new studies have been added since then. 
\par
Ammar et al.\cite{IoTSurveyonSecurityFrameworks} discuss the security architecture, requirements and standards in eight IoT platforms, referring to the hardware compatibility and software requirements. 
\par
ML techniques to detect security intrusion in IoT networks were identified in a survey on machine learning (ML) intrusion detection approaches\cite{iotMLSurvey}. It discusses how these techniques can improve IoT security, and presents some issues of those techniques and how those issues can be solved.
\par
Another related survey considers the healthcare system \cite{iotTechnologiesInHealthcareSurvey}; it explores the security and privacy issues in healthcare IoT systems built on the Wireless Body Area Network (WBAN). The survey describes the network topologies and security features and highlights the quality of service and real-time issues in health monitoring systems.  \par
Celik et al. \cite{PACommoditySurvey} consider exploring privacy and security issues in IoT and the program-analysis techniques that help identify security threats against IoT systems. The study discusses the main problems challenging IoT security and privacy and how these could be tackled, considering the program analysis design and implementation for ensuring IoT security.
A security Considerations survey was conducted by Jurcut et al.\cite{SecurityConsiderationSurvey}, describing general security threats and attacks against IoT devices, how they can be prevented, and possible solutions. The survey mentions its contribution to the safety issues, but it was not well identified. The study discusses risk mitigation and prevention associated with security breaches and how they can be improved. \par
In the IoT conflict comparison domain, Ibrahim et al. identified their primary contribution as finding out the conflict's classifications in recent IoT research work and the detection techniques for the identified conflicts\cite{ConflictsClassificationSurvey}. The survey work focuses on security and safety violations in IoT systems, considering the program analysis techniques and techniques to improve the results. On the other hand, some recent tools are not explored in this work as these tools are just recently published. \par
In addition to security and privacy, other issues were explored, e.g., Legal accountability \cite{IoTApplicationsSecurityPrivacySurvey}. The paper highlights the main shortage in the domain, e.g., Universal standardization and protocols. It also suggests providing technologies for interoperability could contribute to solving IoT issues. \par
 A survey in industrial frameworks for the internet of things  \cite{industrialIoTSurvey} explores seven Industrial IoT (IIoT) frameworks and compares them in terms of the provided features and capabilities.  The discussed frameworks are Arrowhead, AutoSAR, BaSys, FiWare, Industrial Data Space (IDS), Open Connectivity Foundation (OCF) and IoTivity, Open Mobile Alliance (OMA) SpecWorks-LWM2M, and Potential Frameworks and Standards. The study provides a comparison based on industrial requirements. The main features discussed in this study are Entry to the market Barriers, Interoperability, Security, and Functional features (including Real-Time Features, Runtime features, Centralized or Distributed Approach, Hardware Requirements, Quality of Service.).

\section{Conclusion}
IoT systems are overgrowing, which results in more connected IoT devices, more developed IoT apps, and more deployed IoT platforms. This growth introduces new challenges to the IoT systems and their environments. One challenge could be the interactions among different IoT apps developed using different IoT platforms and could run together in one environment. Some motivating example highlights the safety impact of IoT system on its users or environment due to a security breach. 
This study surveys the recent approaches developed to tackle this problem or contribute to it. The study highlights the main analysis techniques used in the surveyed tools, the utilization of IoT security and safety verification when conducting the analysis, and whether the studied tools consider the surrounding IoT environment. 
\begin{acknowledgements}
This work is supported in part by the Natural Sciences and Engineering Research Council of Canada (NSERC), Grant No.RGPIN/06283-2018
\end{acknowledgements}

 \section*{Conflict of interest}
 The authors declare that they have no conflict of interest.

\bibliographystyle{unsrturl}
\bibliography{ref.bib}

\begin{thebibliography}{10}

\bibitem{towardIoT}
Roberto Minerva, Abyi Biru, and Domenico Rotondi.
\newblock Towards a definition of the internet of things (iot).
\newblock {\em IEEE Internet Initiative}, 1(1):1--86, 2015.

\bibitem{Statista}
Statista.
\newblock Iot and non-iot connections worldwide 2010-2025, 2021.
\newblock Last accessed 17 October 2021.
\newblock URL:
  \url{https://www.statista.com/statistics/1101442/iot-number-of-connected-devices-worldwide/}.

\bibitem{Statista2}
Statista.
\newblock Internet-connected light-duty vehicles sales worldwide and in the
  united states in 2023, 2021.
\newblock Last accessed 18 October 2021.
\newblock URL:
  \url{https://www.statista.com/statistics/275849/number-of-vehicles-connected-to-the-internet/}.

\bibitem{iotanalytics}
Statista.
\newblock Number of connected iot devices growing 9\% to 12.3 billion globally,
  2021.
\newblock Last accessed 20 October 2021.
\newblock URL: \url{{}https://iot-analytics.com/number-connected-iot-devices/}.

\bibitem{SmartThings}
Samsung SmartThings.
\newblock One simple home system. a world of possibilities., 2021.
\newblock Last accessed 20 December 2021.
\newblock URL: \url{https://www.smartthings.com/}.

\bibitem{OpenHAB}
OpenHAB.
\newblock Empowering the smart home, 2021.
\newblock Last accessed 20 December 2021.
\newblock URL: \url{https://www.openhab.org/}.

\bibitem{IFTTT}
IFTTT.
\newblock Every thing works better together, 2021.
\newblock Last accessed 20 December 2021.
\newblock URL: \url{https://ifttt.com/}.

\bibitem{PACommoditySurvey}
Z~Berkay Celik, Earlence Fernandes, Eric Pauley, Gang Tan, and Patrick
  McDaniel.
\newblock {Program Analysis of Commodity IoT applications for security and
  privacy: Challenges and opportunities}.
\newblock {\em ACM Computing Surveys (CSUR)}, 52(4):1--30, 2019.

\bibitem{IoTPlatformsSecurityPrivacySurvey}
Leonardo Babun, Kyle Denney, Z~Berkay Celik, Patrick McDaniel, and A~Selcuk
  Uluagac.
\newblock {A survey on IoT platforms: Communication, security, and privacy
  perspectives}.
\newblock {\em Computer Networks}, 192:108040, 2021.

\bibitem{IoTApplicationsSecurityPrivacySurvey}
Parul Goyal, Ashok~Kumar Sahoo, Tarun~Kumar Sharma, and Pramod~K Singh.
\newblock {Internet of Things: Applications, security and privacy: A survey}.
\newblock {\em Materials Today: Proceedings}, 34:752--759, 2021.

\bibitem{SecurityConsiderationSurvey}
Anca Jurcut, Tiberiu Niculcea, Pasika Ranaweera, and Nhien-An Le-Khac.
\newblock {Security considerations for Internet of Things: A survey}.
\newblock {\em SN Computer Science}, 1:1--19, 2020.

\bibitem{IoTSurveyonSecurityFrameworks}
Mahmoud Ammar, Giovanni Russello, and Bruno Crispo.
\newblock {Internet of Things: A survey on the security of IoT frameworks}.
\newblock {\em Journal of Information Security and Applications}, 38:8--27,
  2018.

\bibitem{IoTSecurityServey}
Alma Oracevic, Selma Dilek, and Suat Ozdemir.
\newblock Security in internet of things: A survey.
\newblock In {\em 2017 International Symposium on Networks, Computers and
  Communications (ISNCC)}, pages 1--6. IEEE, 2017.

\bibitem{industrialIoTSurvey}
Cristina Paniagua and Jerker Delsing.
\newblock Industrial frameworks for internet of things: A survey.
\newblock {\em IEEE Systems Journal}, 15(1):1149--1159, 2020.

\bibitem{iotTechnologiesInHealthcareSurvey}
Mrinai~M Dhanvijay and Shailaja~C Patil.
\newblock Internet of things: A survey of enabling technologies in healthcare
  and its applications.
\newblock {\em Computer Networks}, 153:113--131, 2019.

\bibitem{ConflictsClassificationSurvey}
Hamada Ibrhim, Hesham Hassan, and Emad Nabil.
\newblock {A conflicts’ classification for IoT-based services: a comparative
  survey}.
\newblock {\em PeerJ Computer Science}, 7:e480, 2021.

\bibitem{ieeexplore}
{IEEEXplore}, 2022.
\newblock Last accessed 12 Jan 2022.
\newblock URL: \url{https://ieeexplore.ieee.org/}.

\bibitem{ACM}
ACM.
\newblock The association for computing machinery, 2022.
\newblock Last accessed 12 Jan 2022.
\newblock URL: \url{https://dl.acm.org/}.

\bibitem{ScienceDirect}
{ScienceDirect}, 2022.
\newblock Last accessed 12 Jan 2022.
\newblock URL: \url{https://www.sciencedirect.com/}.

\bibitem{Usenix}
Usenix the advanced computing systems association, 2022.
\newblock Last accessed 20 Jan 2022.
\newblock URL: \url{https://www.usenix.org/}.

\bibitem{GoogleScholar}
{GoogleScholar}, 2022.
\newblock Last accessed 12 Jan 2022.
\newblock URL: \url{https://scholar.google.com/}.

\bibitem{HomeKit}
HomeKit.
\newblock ios - home- apple, 2021.
\newblock Last accessed 20 December 2021.
\newblock URL: \url{https://www.apple.com/ios/home/}.

\bibitem{HomeAssistant}
Home Assistant~Core Team and Community.
\newblock Home assistant, 2021.
\newblock Last accessed 20 December 2021.
\newblock URL: \url{https://www.home-assistant.io/}.

\bibitem{AmazonAws}
Amazon AWS.
\newblock Free cloud computing services, 2021.
\newblock Last accessed 20 December 2021.
\newblock URL: \url{https://aws.amazon.com/}.

\bibitem{EvaICS}
Altertech Group.
\newblock {EVA ICS} smart automation platform. industrial and home iot., 2021.
\newblock Last accessed 20 December 2021.
\newblock URL: \url{https://www.eva-ics.com/}.

\bibitem{Patriot}
Moosa Yahyazadeh, Syed~Rafiul Hussain, Endadul Hoque, and Omar Chowdhury.
\newblock {PatrIoT: Policy Assisted Resilient Programmable IoT System}.
\newblock In {\em International Conference on Runtime Verification}, pages
  151--171. Springer, 2020.

\bibitem{WeConIoT}
Sai~Bhatt Keshipeddi.
\newblock Iot based smart water quality monitoring system.
\newblock {\em Available at SSRN 3904842}, 2021.

\bibitem{healthSaaS}
Cornel Turcu and Cristina Turcu.
\newblock Improving the quality of healthcare through internet of things.
\newblock {\em arXiv preprint arXiv:1903.05221}, 2019.

\bibitem{tap2021}
Andrea Mattioli and Fabio Patern{\`o}.
\newblock Recommendations for creating trigger-action rules in a block-based
  environment.
\newblock {\em Behaviour \& Information Technology}, pages 1--11, 2021.

\bibitem{tapSafeChain}
Kai-Hsiang Hsu, Yu-Hsi Chiang, and Hsu-Chun Hsiao.
\newblock Safechain: Securing trigger-action programming from attack chains.
\newblock {\em IEEE Transactions on Information Forensics and Security},
  14(10):2607--2622, 2019.

\bibitem{StaticDynamicHybridComparison}
Anusha Damodaran, Fabio~Di Troia, Corrado~Aaron Visaggio, Thomas~H Austin, and
  Mark Stamp.
\newblock A comparison of static, dynamic, and hybrid analysis for malware
  detection.
\newblock {\em Journal of Computer Virology and Hacking Techniques},
  13(1):1--12, 2017.

\bibitem{DynamicAnalysisSurvey}
Manuel Egele, Theodoor Scholte, Engin Kirda, and Christopher Kruegel.
\newblock A survey on automated dynamic malware-analysis techniques and tools.
\newblock {\em ACM computing surveys (CSUR)}, 44(2):1--42, 2008.

\bibitem{SMTSolvers}
Nikita Malyshev, Irina Dudina, Daniil Kutz, Alexander Novikov, and Sergey
  Vartanov.
\newblock {SMT Solvers in Application to Static and Dynamic Symbolic Execution:
  A Case Study}.
\newblock In {\em 2019 Ivannikov Ispras Open Conference (ISPRAS)}, pages 9--15.
  IEEE, 2019.

\bibitem{modelChecking}
Ranjit Jhala and Rupak Majumdar.
\newblock Software model checking.
\newblock {\em ACM Computing Surveys (CSUR)}, 41(4):1--54, 2009.

\bibitem{symbolicExecutionSurvey}
Roberto Baldoni, Emilio Coppa, Daniele~Cono D’elia, Camil Demetrescu, and
  Irene Finocchi.
\newblock A survey of symbolic execution techniques.
\newblock {\em ACM Computing Surveys (CSUR)}, 51(3):1--39, 2018.

\bibitem{SecuringIoT}
Sridipta Misra, Muthucumaru Maheswaran, and Salman Hashmi.
\newblock Securing the internet of things.
\newblock In {\em Security challenges and approaches in Internet of Things},
  pages 39--51. Springer, 2017.

\bibitem{SecuringIoTRoadAhead}
Sabrina Sicari, Alessandra Rizzardi, Luigi~Alfredo Grieco, and Alberto
  Coen-Porisini.
\newblock Security, privacy and trust in internet of things: The road ahead.
\newblock {\em Computer networks}, 76:146--164, 2015.

\bibitem{IoTSafetyStateofArt}
Janusz Zalewski.
\newblock Iot safety: state of the art.
\newblock {\em IT Professional}, 21(1):16--20, 2019.

\bibitem{JeepCherooke}
Wired.
\newblock Hackers remotely kill a jeep on the highway—with me in it, 2015.
\newblock Last accessed 21 October 2021.
\newblock URL:
  \url{https://www.wired.com/2015/07/hackers-remotely-kill-jeep-highway/}.

\bibitem{Soteria}
Z~Berkay Celik, Patrick McDaniel, and Gang Tan.
\newblock Soteria: Automated {IoT} safety and security analysis.
\newblock In {\em {2018 USENIX Annual Technical Conference}}, pages 147--158,
  2018.

\bibitem{Iotsan}
Dang~Tu Nguyen, Chengyu Song, Zhiyun Qian, Srikanth~V Krishnamurthy, Edward~JM
  Colbert, and Patrick McDaniel.
\newblock {IoTSan}: Fortifying the safety of {IoT} systems.
\newblock In {\em Proceedings of the 14th International Conference on emerging
  Networking EXperiments and Technologies}, pages 191--203, 2018.

\bibitem{IoTMoNn}
Wenbo Ding and Hongxin Hu.
\newblock On the safety of {IoT} device physical interaction control.
\newblock In {\em Proceedings of the 2018 ACM SIGSAC Conference on Computer and
  Communications Security}, pages 832--846, 2018.

\bibitem{ProvThings}
Qi~Wang, Wajih~Ul Hassan, Adam Bates, and Carl Gunter.
\newblock Fear and logging in the internet of things.
\newblock In {\em Network and Distributed Systems Symposium}, 2018.

\bibitem{Foresee}
Zheng Fang, Hao Fu, Tianbo Gu, Zhiyun Qian, Trent Jaeger, and Prasant
  Mohapatra.
\newblock Foresee: A cross-layer vulnerability detection framework for the
  internet of things.
\newblock In {\em 2019 IEEE 16th International Conference on Mobile Ad Hoc and
  Sensor Systems (MASS)}, pages 236--244. IEEE, 2019.

\bibitem{viscr}
Vasudevan Nagendra, Arani Bhattacharya, Vinod Yegneswaran, Amir Rahmati, and
  Samir~R Das.
\newblock {VISCR}: intuitive \& conflict-free automation for securing the
  dynamic consumer iot infrastructures.
\newblock {\em arXiv preprint arXiv:1907.13288}, 2019.

\bibitem{HomeGuard}
Haotian Chi, Qiang Zeng, Xiaojiang Du, and Jiaping Yu.
\newblock Cross-app interference threats in smart homes: Categorization,
  detection and handling.
\newblock In {\em 2020 50th Annual IEEE/IFIP International Conference on
  Dependable Systems and Networks (DSN)}, pages 411--423. IEEE, 2020.

\bibitem{IoTcomm}
Mohannad Alhanahnah, Clay Stevens, and Hamid Bagheri.
\newblock Scalable analysis of interaction threats in {IoT} systems.
\newblock In {\em Proceedings of the 29th ACM SIGSOFT international symposium
  on software testing and analysis}, pages 272--285, 2020.

\bibitem{iruler}
Qi~Wang, Pubali Datta, Wei Yang, Si~Liu, Adam Bates, and Carl~A Gunter.
\newblock Charting the attack surface of trigger-action {IoT} platforms.
\newblock In {\em Proceedings of the 2019 ACM SIGSAC conference on computer and
  communications security}, pages 1439--1453, 2019.

\bibitem{Remediot}
Renju Liu, Ziqi Wang, Luis Garcia, and Mani Srivastava.
\newblock {RemedIoT}: Remedial actions for internet-of-things conflicts.
\newblock In {\em Proceedings of the 6th ACM International Conference on
  Systems for Energy-Efficient Buildings, Cities, and Transportation}, pages
  101--110, 2019.

\bibitem{Autotap}
Lefan Zhang, Weijia He, Jesse Martinez, Noah Brackenbury, Shan Lu, and Blase
  Ur.
\newblock {AutoTap}: synthesizing and repairing trigger-action programs using
  {LTL} properties.
\newblock In {\em 2019 IEEE/ACM 41st international conference on software
  engineering (ICSE)}, pages 281--291. IEEE, 2019.

\bibitem{infoFlowModel}
Milijana Surbatovich, Jassim Aljuraidan, Lujo Bauer, Anupam Das, and Limin Jia.
\newblock Some recipes can do more than spoil your appetite: Analyzing the
  security and privacy risks of {IFTTT} recipes.
\newblock In {\em Proceedings of the 26th International Conference on World
  Wide Web}, pages 1501--1510, 2017.

\bibitem{Safetap}
McKenna McCall, Faysal~Hossain Shezan, Abhishek Bichhawat, Camille Cobb, Limin
  Jia, Yuan Tian, Cooper Grace, and Mitchell Yang.
\newblock Safetap: An efficient incremental analyzer for trigger-action
  programs.
\newblock {\em Carnegie Mellon University}, 2021.

\bibitem{Pfirewall}
Haotian Chi, Qiang Zeng, Xiaojiang Du, and Lannan Luo.
\newblock {PFirewall: Semantics-aware customizable data flow control for home
  automation systems}.
\newblock {\em arXiv preprint arXiv:1910.07987}, 2019.

\bibitem{Peeves}
Simon Birnbach, Simon Eberz, and Ivan Martinovic.
\newblock Peeves: Physical event verification in smart homes.
\newblock In {\em Proceedings of the 2019 ACM SIGSAC Conference on Computer and
  Communications Security}, pages 1455--1467, 2019.

\bibitem{Iotguard}
Z~Berkay Celik, Gang Tan, and Patrick~D McDaniel.
\newblock {IoTGuard: Dynamic Enforcement of Security and Safety Policy in
  Commodity IoT}.
\newblock In {\em NDSS}, 2019.

\bibitem{IoTBox}
Hong~Jin Kang, Sheng~Qin Sim, and David Lo.
\newblock {IoTBox: Sandbox Mining to Prevent Interaction Threats in IoT
  Systems}.
\newblock In {\em 2021 14th IEEE Conference on Software Testing, Verification
  and Validation (ICST)}, pages 182--193. IEEE, 2021.

\bibitem{Smartattack}
Keyang Yu and Dong Chen.
\newblock {SmartAttack: Open-source Attack Models for Enabling Security
  Research in Smart Homes}.
\newblock In {\em 2020 11th International Green and Sustainable Computing
  Workshops (IGSC)}, pages 1--8. IEEE, 2020.

\bibitem{IoTPraetor}
Juan Wang, Shirong Hao, Ru~Wen, Boxian Zhang, Liqiang Zhang, Hongxin Hu, and
  Rongxing Lu.
\newblock {IoT-Praetor: Undesired behaviors detection for IoT devices}.
\newblock {\em IEEE Internet of Things Journal}, 8(2):927--940, 2020.

\bibitem{HAWatcher}
Chenglong Fu, Qiang Zeng, and Xiaojiang Du.
\newblock $\{$HAWatcher$\}$:$\{$Semantics-Aware$\}$ anomaly detection for
  appified smart homes.
\newblock In {\em 30th USENIX Security Symposium (USENIX Security 21)}, pages
  4223--4240, 2021.

\bibitem{MenShen}
Lei Bu, Wen Xiong, Chieh-Jan~Mike Liang, Shi Han, Dongmei Zhang, Shan Lin, and
  Xuandong Li.
\newblock {Systematically ensuring the confidence of real-time home automation
  IoT systems}.
\newblock {\em ACM Transactions on Cyber-Physical Systems}, 2(3):1--23, 2018.

\bibitem{SOK}
Omar Alrawi, Chaz Lever, Manos Antonakakis, and Fabian Monrose.
\newblock {SoK: Security Evaluation of Home-based IoT Deployments}.
\newblock In {\em 2019 IEEE symposium on security and privacy (sp)}, pages
  1362--1380. IEEE, 2019.

\bibitem{VETPLC}
Mu~Zhang, Chien-Ying Chen, Bin-Chou Kao, Yassine Qamsane, Yuru Shao, Yikai Lin,
  Elaine Shi, Sibin Mohan, Kira Barton, James Moyne, et~al.
\newblock {Towards Automated Safety Vetting of PLC Code in real-world plants}.
\newblock In {\em 2019 IEEE Symposium on Security and Privacy (SP)}, pages
  522--538. IEEE, 2019.

\bibitem{IotGaze}
Tianbo Gu, Zheng Fang, Allaukik Abhishek, Hao Fu, Pengfei Hu, and Prasant
  Mohapatra.
\newblock {IoTGaze: IoT Security enforcement via wireless context analysis}.
\newblock In {\em IEEE INFOCOM 2020-IEEE Conference on Computer
  Communications}, pages 884--893, 2020.

\bibitem{SmartMon}
Pengfei Peng and An~Wang.
\newblock {SmartMon: Misbehavior Detection via Monitoring Smart Home
  Automations}.
\newblock In {\em 2020 IEEE/ACM Symposium on Edge Computing (SEC)}, pages
  327--333. IEEE, 2020.

\bibitem{SmartFuzz}
Lwin~Khin Shar, Ta~Nguyen~Binh Duong, Lingxiao Jiang, David Lo, Wei Minn, Glenn
  Kiah~Yong Yeo, and Eugene Kim.
\newblock {SmartFuzz: An Automated Smart Fuzzing Approach for Testing
  SmartThings Apps}.
\newblock In {\em 2020 27th Asia-Pacific Software Engineering Conference
  (APSEC)}, pages 365--374. IEEE, 2020.

\bibitem{Jarvis}
Anand Mudgerikar and Elisa Bertino.
\newblock {Jarvis: Moving Towards a Smarter Internet of Things}.
\newblock In {\em 2020 IEEE 40th International Conference on Distributed
  Computing Systems (ICDCS)}, pages 122--134. IEEE, 2020.

\bibitem{Iotsafe}
Wenbo Ding, Hongxin Hu, and Long Cheng.
\newblock Iotsafe: Enforcing safety and security policy with real iot physical
  interaction discovery.
\newblock In {\em the 28th Network and Distributed System Security Symposium
  (NDSS 2021)}, 2021.

\bibitem{IoTSEER}
Muslum~Ozgur Ozmen, Xuansong Li, Andrew Chun-An Chu, Z~Berkay Celik, Bardh
  Hoxha, and Xiangyu Zhang.
\newblock {Discovering Physical Interaction Vulnerabilities in IoT
  Deployments}.
\newblock {\em arXiv preprint arXiv:2102.01812}, 2021.

\bibitem{Hybridiagnostics}
Abhinav Mohanty and Meera Sridhar.
\newblock {HybriDiagnostics: Evaluating Security Issues in Hybrid SmartHome
  Companion Apps}.
\newblock In {\em 2021 IEEE Security and Privacy Workshops (SPW)}, pages
  228--234. IEEE, 2021.

\bibitem{Expat}
Moosa Yahyazadeh, Proyash Podder, Endadul Hoque, and Omar Chowdhury.
\newblock {Expat: Expectation-Based Policy Analysis and Enforcement for
  Appified smart-home platforms}.
\newblock In {\em Proceedings of the 24th ACM Symposium on Access Control
  Models and Technologies}, pages 61--72, 2019.

\bibitem{Tapinspector}
Yinbo Yu and Jiajia Liu.
\newblock {TAPInspector: Safety and Liveness Verification of Concurrent
  Trigger-Action IoT Systems}.
\newblock {\em arXiv preprint arXiv:2102.01468}, 2021.

\bibitem{FriendlyFire}
Michele~Pasqua Musard~Balliu, Massimo~Merro.
\newblock Securing cross-app interactions in iot platforms.
\newblock In {\em 2019 IEEE 32nd Computer Security Foundations Symposium
  (CSF)}, page 319–334, 2019.

\bibitem{stpa}
Takuo Hayakawa, Ryoichi Sasaki, Hiroshi Hayashi, Yuji Takahashi, Tomoko Kaneko,
  and Takao Okubo.
\newblock Proposal and application of security/safety evaluation method for
  medical device system that includes iot.
\newblock In {\em Proceedings of the 2018 VII International Conference on
  Network, Communication and Computing}, pages 157--164, 2018.

\bibitem{Savior}
Raul Quinonez, Jairo Giraldo, Luis Salazar, Erick Bauman, Alvaro Cardenas, and
  Zhiqiang Lin.
\newblock {SAVIOR: Securing Autonomous Vehicles with Robust Physical
  Invariants}.
\newblock In {\em 29th USENIX Security Symposium USENIX Security 20}, pages
  895--912, 2020.

\bibitem{iotMLSurvey}
Kelton~AP da~Costa, Jo{\~a}o~P Papa, Celso~O Lisboa, Roberto Munoz, and Victor
  Hugo~C de~Albuquerque.
\newblock {Internet of Things: A survey on machine learning-based intrusion
  detection approaches}.
\newblock {\em Computer Networks}, 151:147--157, 2019.

\end{thebibliography}
\end{document}